\documentclass{icrc2009}

\usepackage{graphicx}   % for including figures
\usepackage{caption}    % for captions
\usepackage[font=footnotesize]{subfig} % subfig.sty for a double column floating figure using two subfigures
\usepackage{fixltx2e}
\usepackage{url}

\newcommand{\shorttitle}[1]%
{\markboth{Proceedings of the 31\MakeLowercase{$^{st}$} ICRC, {\L}\'{o}d\'{z} 2009}{#1} }
\newcommand{\etal}{\MakeLowercase{\textit{et al. }}} % "et al."

\usepackage{amsmath}
\usepackage{amsfonts}
\usepackage{amssymb}

\begin{document}
\title{Dark Energy and Search for the Generalised Second Law}

\author{Balendra Kr. Dev Choudhury \\
Julie Saikia \\ 
Deptt. of Physics, Pub Kamrup College\\
Baihata Chariali - 781381, Assam (INDIA)}
\date{}

% please write the preseter's name and short title (3-4 words maximum)
%    which will appear at the header of the even pages.
\shorttitle{ Dev Choudhury \etal Search for the GSL}
\maketitle

\begin{abstract}
\noindent The discovery of accelerated Hubble expansion in the SN Ia data and the observed power spectrum of the microwave background radiation provide an ample support for Dark energy and Dark matter. Except for the so far well - known facts that cold dark matter (or simply dark matter) is pressureless, and dark energy has a negative pressure, the nature of these two remains a complete mystery. The mystery facilitates different consideration. In one hand, dark matter and dark energy are assumed as distinct entities, and other interpretation is that both are different manifestation of a common structure, often referred as quartessence. Chaplygin gas, a perfect fluid also favours the second interpretation. Here, we consider modified chaplygin gas as dark energy candidate. Taking into account the existence of the observer's event horizon in accelerated universe, we find the condition where the generalised second law of gravitational thermodynamics is valid and the positivity of the temperature of the phantom fluid remains intact.
\end{abstract}

 
\section{Introduction}

\noindent On the basis of the observational data [1], cosmologists are in the firm ground to look beyond standard cosmology. Instead of predicted deceleration in the present universe by standard cosmology, they have to consider the late cosmic acceleration. The main actor in this scenario is some exotic fluid, and its dominating influence in the very late universe. In Newtonian mechanics, such type of exotic fluids cannot be formulated. Its characteristic negative pressure $p < -\left( \frac{1}{3}\right)\rho$ with $\rho$ being the energy density, can be obtained from Freidmann equations. In a very naive way, one can realise that since in the general relativistic picture, gravitation is thought to be the consequence of space-time curvature, differently curved space-time can generate also negative pressure.\\

\noindent Negative pressure requires the violation of the strong energy condition (SEC), $\rho + 3p > 0$. The very exotic fluid takes its popular name dark energy (DE) due to its unknown character, and having gravitational effect [2]. On the contrary to that condition, $\rho + 3p < 0$ implying $-\frac{1}{3} > \omega > - 1$ with equation of state parmeter $\omega = \frac{p}{\rho}$, Caldwell showed the case $\omega < -1$ better fit for the observed astrophysical data and preferred this case, which violates the dominant energy condition, $\rho + p > 0$ as well [3]. In general, fluid of such characteristics is known as phantom. In spite of its better data - fitting, phantom shows our incomplete understanding of the physics below $\rho + p = 0$ [4]. One major issue, in this field that warrants further explanation is the thermodynamical behaviour of the fluid. \\

\noindent Negative pressure leading to an accelerated universe can also be the characteristic of chaplygin gas. In addition to its exotic equation of state, it has some interesting features [5]. Chaplygin gas equation of state is connected to string theory via a brane interpretation. Moreover, it is reported that the chaplygin gas is the only gas known to admit a supersymmetric generalisation [6]. In course of the different evolving scenario, chaplygin gas is also accordingly evolved. Modified chaplygin gas is observed to interpolate between standard fluids at high energy densities and chaplygin gas fluids at low energy densities. Its unifying character that unifies dark matter and dark enrgy is really interesting. Modified chaplygin gas is represented by an exotic equation [7]\\
\begin{equation}
p = \gamma \rho - \frac{B}{\rho^\alpha}, \qquad B > 0,  \qquad 0 < \alpha < 1
\end{equation} \\

\noindent In this paper, we consider the issue of generalised second law of gravitational thermodynamics in the context of modified chaplygin gas.

\section {Generalised Second Law and Dark Energy}

\noindent If the expansion of the universe is dominated by phantom fluid, black holes are to loss their mass, and thus, are bound to disappear. As black holes are most entropic bodies in the Universe, its disappearance causes the decrease of entropy of the Universe. In a very elementary way, the answer becomes that it cannot be according to the second law of thermodynamics [4]. But this negative approach has to be abandoned due to firm theoretical ground. Assigning an entropy to the horizon size, the second law has to be generalised. According to GSL, the entropy of matter and fields inside the horizon plus the entropy of the event horizon cannot decrease with time. Thus\\

\begin {equation}
\dot{S} + \dot{S_H} \geq 0
\end {equation}

\noindent Pollock and Singh [8], for the first time, considered this issue assuming the energy sources that violate the dominant energy condition. Extending their work from various points, Izquierdo and Pav$\acute{o}$n give some hints for further works, specially in a less idealised cosmology [9]. We shall follow these points in our work in the context of modified chaplygin gas.\\

\section {Modified Chaplygin Gas and GSL}

\noindent Considering the EOS of modified chaplygin gas (1), and incorporating the conservation equation for the homogeneous perfect fluid constituting the cosmic fluid

$$\dot{\rho} + 3 H \left( \rho + p\right) = 0$$\\

\noindent we have

$$ \rho ^ {\alpha + 1} = \frac{B} {1 + \gamma} + \frac{C} {a^ {3\left( 1 + \alpha\right) \left( 1 + \gamma\right)}}$$\\

\noindent where $C$ is the constant of integration.\\

\noindent In case of phantom fluid

$$p = \gamma \rho - \frac{B}{\rho ^ \alpha} < -\rho $$\\

\noindent Now, let us suppose\\

$$ \omega \left( = \frac{p}{\rho}\right) \equiv \omega_0 = \gamma - \frac{B}{\rho_0 ^{\left( \alpha + 1\right)}}= -1.06 $$\\

\noindent Then it comes

$$\rho_0 ^ {\alpha + 1} =  \frac{B \rho_0 ^ {\left( \alpha + 1\right) }} {B - 0.06 \rho_0 ^{\alpha + 1}} + \frac{C}{a_0 ^ {3 \left( \alpha + 1\right)\left( 1 + \gamma\right) } }$$\\

\noindent Setting $a_0 = 1$, we have

$$ C = \frac{- 0.06 \rho_0 ^ {2 \left( \alpha + 1\right) }} {B - 0.06 \rho_0 ^ {\alpha + 1}}$$\\

\noindent Let

$$ B = n \rho_0 ^ {\alpha + 1}$$\\

\noindent Considering $n = 0.56$, we get $ B = - 0.5 $\\

\noindent Then, we may obtain

$$ \rho ^ {\alpha + 1} = 2 \left[ 0.56 - \frac{0.06}{a ^ {\frac {3\left( \alpha + 1\right)} {2}}}\right] \rho+0 ^ {\alpha + 1}$$\\

\noindent Now using Friedmann equation, we can have

\begin{equation}
\left( \frac{\dot{a}}{a}\right) ^ 2  = \frac{8 \pi}{3 M_P ^ 2} \left( 2\right) ^ {\frac{1}{\alpha + 1}}
\rho_ 0 \frac {\left[ 0 . 56 a ^ {\frac{3 \left( \alpha + 1 \right)} {2}} - 0.06 \right] ^ {\frac{1}{\alpha + 1}}} {a ^ {\frac{3}{2}}}
\end{equation}\\

\noindent And ultimately, the scale factor reads

$$a = [ \frac{3}{28} + \frac{25}{14}\lbrace \left( 0.56a_*^{3(1+\alpha)/2} - 0.06\right)  ^{\frac{1}{\alpha + 1}}$$\\

\begin{equation}
+ A(t - t_*) \rbrace  ^{(1+\alpha)}] ^{\frac{2} {3(1+\alpha)}}
\end{equation}\\

\noindent where $A^2=\frac{9}{4} \left( \frac{8\pi} {3 {M_p}^2} (2)^{\frac{1}{\alpha+1}} \rho_0 \right)$ \\

\noindent The expression of the radius of the observer's event horizon reads

$$ R_H = a(t)\int _{t_*}^t \frac{dt} {a(t)}$$\\

\noindent And, here the expression for $R_H$ comes out as

\begin{equation}
R_H \cong \frac{3 a (t)}{A} \left[ a ^ \frac{1}{2} - \left\lbrace 0.56 \left( \frac{25}{14}\right) \right\rbrace ^ \frac{1}{2} a_* ^ \frac{1}{2}\right]
\end{equation}\\

\noindent We consider the phantom fluid inside the cosmological event horizon of a comoving observer. The relation of the entropy with energy and pressure of the fluid can be expressed in terms of Gibb's equation

\begin{equation}
T ds = d E + P d \left( \frac{4}{3} \pi R_ H ^3\right) 
\end{equation}

\noindent After some simplification, the expression becomes

\begin{equation}
T ds = - \frac{4 \pi R_H ^2}{4 \pi G} \dot{H} \left( \frac{3}{2 A}\right)  a ^ {\frac{1}{2}}
\end{equation}\\

\noindent If the temperature of the event horizon is assumed as the temperature of the fluid i.e, $T = T_H = \frac{1}{2\pi R_H}$, we have

\begin{equation}
 ds = - \frac{2 \pi R_H ^3}{G} \dot{H} \left( \frac{3}{2 A}\right)  a ^ {\frac{1}{2}}
\end{equation}\\

\section{Discussion}

\noindent Izquierdo and Pav$\acute{o}$n observed that in a less idealised cosmology, the presence of other forms of energy is to be considered. The validity of GSL is also not clear. Among other things, the expression for the scale factor also should not be a simple one like $a(t)\propto \frac{1}{(t_* - t) ^n}$,  $\left( t \leq {t_*},  0 < n = const\right)$. Here, we have obtained the scale factor (3) following the whole dynamics and the equation of state of modified chaplygin gas, and really it is an involved expression. And for our concerned temperature, we have found from the expression (7) that phantom fluids, at least for $\omega = -1.06$, possess negative entropy if $\dot{H} > 0$, and the entropy is positive if $\dot{H} < 0$. The expression also clearly shows the change of entropy is parabolic in nature.\\

\noindent {\bf{Acknowledgements}} \\

\noindent Late Prof. S. K. Srivastava, North-Eastern Hill University (India) provided us his guidance and encouragement during this work in spite of his poor health.  We are indebted, and painfully remember him. \\

\end{document}